\documentclass[aps,prd,10pt,notitlepage,nofootinbib,superscriptaddress,showkeys,showpacs]{revtex4-2}
\pdfoutput=1
\usepackage{amstext,amsmath,amssymb,amsfonts,bbm,euscript,color,dsfont}
\usepackage[utf8]{inputenc}
\usepackage{epsfig}
\usepackage{hyperref}
\usepackage{slashed}
\usepackage{amsthm}
\usepackage{subfigure}
\usepackage{xcolor}
\usepackage{multirow}
\usepackage{bbold}

\usepackage{bm}
\usepackage{tikz}
\usetikzlibrary{positioning,arrows.meta}
\usepackage{verbatim}
\usepackage{graphicx}
\usepackage{latexsym}

\begin{document}

\title{Glueball mass from RGZ-inspired infrared gluodynamics: a Euclidean Bethe–Salpeter approach}

\author{Rodrigo Carmo Terin} 
\email{rodrigo.carmo@urjc.es}
\affiliation{King Juan Carlos University, Faculty of Experimental Sciences and Technology, Department of Applied Physics, Av. del Alcalde de Móstoles, 28933, Madrid, Spain}

\begin{abstract}
We formulate and solve a Euclidean Bethe-Salpeter equation (BSE) for the lightest $0^{++}$ glueball in pure Yang-Mills (YM) theory, using the the refined Gribov-Zwanziger (RGZ) gluon tree-level propagator as an infrared-complete input.
In a minimal ladder truncation with an effective constant kernel strength $g_C^2$ and the dominant $s$--wave component, we extract scalar glueball masses in the range $M_{0^{++}}\simeq 1.7$--$2.3~\mathrm{GeV}$ for representative values of $g_C^2$, with a preferred value $M_{0^{++}}\simeq 1.9~\mathrm{GeV}$ around $g_C^2\simeq 0.54$.
The result is consistent with RGZ correlator-based infrared moment analyses and with lattice expectations, providing a cross-check of RGZ-inspired infrared gluodynamics from a bound-state viewpoint.
\end{abstract}

\maketitle

\section{Introduction}
\label{sec:introduction}

Understanding the emergence of a physical, colorless spectrum from nonabelian YM theory~\cite{YangMills1954}
remains an important problem in quantum chromodynamics (QCD).
A paradigmatic manifestation of confinement is the expected existence of glueballs, i.e.\ bound states
with purely gluonic valence content.
Although lattice simulations provide robust evidence for a discrete glueball spectrum in pure gauge
theories, connecting these results to continuum frameworks that remain analytically controllable in the
infrared (IR) is still a highly nontrivial task.

In continuum formulations, a gauge choice is required, and the standard Faddeev--Popov (FP)~ \cite{FaddeevPopov1967} procedure
is incomplete in the IR due to the presence of Gribov copies~\cite{Gribov1978,Singer1978}.
In the Landau gauge, large-volume lattice simulations have long established a decoupling-type gluon
propagator: the transverse propagator saturates to a finite nonzero value as $p^2\to 0$ while violating
reflection positivity~\cite{CucchieriMendes2008,Bogolubsky2009,Boucaud2012,Maas2013}.
This behavior strongly suggests that the relevant IR degrees of freedom are not particle-like gluons,
and that physical information must be extracted from gauge-invariant composite operators.

A prominent continuum strategy to incorporate the Gribov problem while preserving locality and
renormalizability is the GZ framework~\cite{Zwanziger1989,Zwanziger1990},
and its Refined version (RGZ), where dimension-two condensates introduce additional IR mass scales
that bring YM correlators in quantitative agreement with lattice data in the Landau gauge
over a broad momentum window~\cite{Dudal2008RGZ,Dudal2010LatticeFit}.
An especially powerful formulation of RGZ employs the gauge-invariant transverse field $A_\mu^h$,
which enables one to retain an exact nilpotent BRST symmetry (in particular in linear covariant gauges),
thereby ensuring a controlled renormalization of gauge-invariant composite operators and the associated
Ward/Nielsen identities~\cite{Sorella2015Nielsen,Sorella2017BRSTinvariant,Terin2021FermionicHorizon}.
The resulting RGZ gluon propagator is IR-finite and typically exhibits complex conjugate poles,
a concise analytic imprint of positivity violation and confinement.

A complementary continuum route, conceptually closer to lattice gauge fixing, is based on lifting the
Gribov degeneracy by averaging over Gribov copies with a tunable weight in the Landau gauge.
This was proposed by J. Serreau and M. Tissier and further work collaborators developed in covariant extensions
and replica/superspace formulations~\cite{SerreauTissier2012,SerreauTissierTresmontant2015, ReinosaSerreauTerinTissier2021}.
In this approach, a replica nonlinear sigma-model sector induces a radiatively generated screening mass
for the gluon through a phenomenon akin to symmetry restoration in the two-dimensional nonlinear sigma
model, yielding an IR-safe perturbative description in a finite domain of renormalized parameters
(without an IR Landau pole at one loop)~\cite{ReinosaSerreauTerinTissier2021}.
More recently, our unified Landau-gauge fixing has been proposed that continuously interpolates between the
Serreau--Tissier (ST) copy-averaged formulation and the RGZ restriction to the first Gribov region,
by combining copy averaging with a horizon suppression in a single local, BRST-invariant and
power-counting renormalizable action~\cite{Terin2025UnifiedST_RGZ}.
This unification gives a controlled theory to investigate how IR YM correlators depend on the balance
between soft copy lifting and hard horizon suppression.

Within RGZ, one can access glueball physics through gauge-invariant correlators.
An influential example is the infrared moment problem (a Pad\'e/Hausdorff moment strategy) applied to the
two-point functions of glueball operators using RGZ-inspired IR gluodynamics, which yields estimates
$m_{0^{++}}\!\sim\!2~\text{GeV}$ and the hierarchy $m_{0^{++}}<m_{2^{++}}<m_{0^{-+}}$ already at the lowest
moment order~\cite{DudalGuimaraesSorella2011PRL}.
This correlator-based route is particularly natural in Euclidean signature, where glueball masses are
extracted from the long-distance decay of Euclidean correlators, as in lattice computations.

Another standard continuum route to bound states is through the famous Bethe--Salpeter equations (BSEs) ~ \cite{BetheSalpeter1951} and related
functional methods (Dyson--Schwinger equations (DSEs) and the functional renormalization group (FRG)),
which organize the resummation of two-particle reducible diagrams into an eigenvalue problem whose
discrete solutions determine the spectrum~\cite{AlkoferVonSmekal2001,Huber2020,Pawlowski2007}.
In hadron physics, there has also been substantial progress on solving the BSE directly in Minkowski space
using the Nakanishi integral representation (NIR) and connecting it to light-front dynamics via
light-front projection, providing direct access to timelike observables and light-front wave functions
for dressed constituents~\cite{Nakanishi1971,CarbonellKarmanov2006,CarbonellKarmanov2010,dePaulaFrederico2026}.
While these Minkowski/light-front developments are mainly targeted at quark bound states, they clarify
formal aspects of relativistic bound-state dynamics and motivate systematic connections between
four-dimensional and projected three-dimensional formulations.

Moreover, continuum approaches based on BSEs have also been applied to the
glueball spectrum using functional methods.
Early DSEs/BSEs studies of scalar and pseudoscalar glueballs can
be found in Ref.~\cite{MeyersSwanson2012Glueballs}, whereas more recent developments include
fully covariant two-gluon BSEs with ghost contributions
\cite{SanchisAlepuz2015Glueballs} and parameter-free truncations derived from
three-particle-irreducible effective actions
\cite{HuberFischerSanchisAlepuz2020}.
These studies consistently report scalar glueball masses in the range
$M_{0^{++}}\sim 1.6$--$2.0~\mathrm{GeV}$, furnishing an important benchmark for our
RGZ-inspired BSE framework developed here.

Therefore, in this work we follow an explicitly Euclidean strategy based on glueballs in pure YM:
we formulate and solve a Euclidean BSE for the lightest scalar glueball
($J^{PC}=0^{++}$), using as nonperturbative input the RGZ gluon propagator obtained in the BRST-invariant
$A_\mu^h$ formulation.
The scalar channel is naturally interpolated by the gauge-invariant local operator
$\mathcal{O}_{0^{++}}=\tfrac14 F_{\mu\nu}^aF_{\mu\nu}^a$.
Our BSE is constructed as an eigenvalue equation for the corresponding two-gluon Bethe--Salpeter amplitude
in the color-singlet channel, with a symmetry-preserving ladder kernel as a first controlled truncation.
Because the entire analysis is performed in Euclidean signature, the complex conjugate pole structure of
the RGZ gluon propagator does not obstruct the integral equation on the integration domain; rather, it
encodes confinement at the level of elementary fields, while the physical spectrum emerges only in the
gauge-invariant composite channel.
Besides providing an independent bound-state perspective complementary to correlator/moment methods,
this Euclidean BSE framework offers direct access to the glueball amplitude and a transparent way to
analyze truncation dependence and IR-input uncertainties.
In particular, we can confront our extracted $0^{++}$ mass with RGZ correlator-based estimates
and with lattice expectations, thereby testing the internal consistency of RGZ-inspired IR gluodynamics
as an effective description of the confining YM infrared sector.

The paper is organized as follows.
Section~\ref{sec:RGZ} reviews the BRST-invariant RGZ framework in terms of $A_\mu^h$ and fixes our
conventions.
Section~\ref{sec:operators} discusses glueball operators and the scalar channel.
Section~\ref{sec:BSE} derives the Euclidean BSE for the $0^{++}$ glueball and specifies the truncation.
Section~\ref{sec:numerics} details the numerical strategy and stability tests.
Section~\ref{sec:results} presents results and compares them with RGZ correlator/moment approaches.
We conclude with an outlook on improving the kernel and extending the analysis to other $J^{PC}$ channels.

\section{The Refined Gribov--Zwanziger framework in the BRST-invariant formulation}
\label{sec:RGZ}

We consider pure YM theory formulated in four-dimensional Euclidean space
and quantized in the Landau gauge,
\begin{align}
\partial_\mu A_\mu^a = 0 \, .
\end{align}
At low energies, the standard FP quantization is known to be incomplete due
to the presence of Gribov copies, i.e. gauge-equivalent configurations that satisfy the same
gauge condition~\cite{Gribov1978,Singer1978}.
The RGZ approach gives an effective infrared completion
of YM theory that incorporates these effects and at the same time preserving locality and
renormalizability~\cite{Zwanziger1989,Zwanziger1990,Dudal2008RGZ}.

A particularly powerful formulation of the RGZ theory is attained by expressing the action
in terms of the gauge-invariant transverse field $A_\mu^h$~\cite{Sorella2015Nielsen,Sorella2017BRSTinvariant}.
This field is defined as the gauge transform of $A_\mu$ that minimizes the functional
\begin{align}
\int d^4x \, A_\mu^a A_\mu^a
\end{align}
along the gauge orbit, and it can be constructed order by order as a nonlocal but
gauge-invariant functional of $A_\mu$.
By construction, $A_\mu^h$ is transverse,
\begin{align}
\partial_\mu A_\mu^{h,a} = 0 \, ,
\end{align}
and invariant under infinitesimal gauge transformations.
This formulation enables the RGZ action to be endowed with an exact nilpotent BRST symmetry,
despite the presence of the Gribov horizon~\cite{Sorella2015Nielsen,Sorella2017BRSTinvariant}.
This property is fundamental in our present work, as it ensures that gauge-invariant
composite operators admit a consistent renormalization and can be meaningfully employed
as interpolating fields in bound-state equations.

The RGZ action can be then written schematically as
\begin{align}
S_{\text{RGZ}}
=
S_{\text{YM}}
+
S_{\text{FP}}
+
S_{\text{horizon}}
+
S_{\text{cond}},
\end{align}
where $S_{\text{YM}}$ denotes the YM action,
$S_{\text{FP}}$ the standard FP gauge-fixing term in the Landau gauge,
and $S_{\text{horizon}}$ implements the restriction of the functional integral
to the first Gribov region.

In the local formulation, the horizon term reads
\begin{align}
S_{\text{horizon}}
&=
\int d^4x \,
\Big[
\bar{\varphi}_\mu^{ac} \,
\mathcal{M}^{ab}(A^h) \,
\varphi_\mu^{bc}
-
\bar{\omega}_\mu^{ac} \,
\mathcal{M}^{ab}(A^h) \,
\omega_\mu^{bc}
\Big]
\nonumber\\
&\quad
-
\gamma^2 g \int d^4x \,
 f^{abc}
A_{\mu}^{h,a}
\left(
\varphi_\mu^{bc}
+
\bar{\varphi}_\mu^{bc}
\right),
\label{eq:horizon}
\end{align}
in which $\mathcal{M}^{ab}(A^h) = -\partial_\mu D_\mu^{ab}(A^h)$ is the FP operator,
$\{\varphi,\bar{\varphi}\}$ are bosonic auxiliary fields,
$\{\omega,\bar{\omega}\}$ their fermionic counterparts,
and $\gamma$ is the Gribov parameter fixed self-consistently by the horizon condition
following from the minimization of the vacuum energy~\cite{Zwanziger1989,Zwanziger1990}.

Nonperturbative effects beyond the horizon restriction are encoded through
dimension-two condensates.
Including their contribution leads to the RGZ action~\cite{Dudal2008RGZ,Dudal2010LatticeFit},
\begin{align}
S_{\text{cond}}
=
\frac{m^2}{2}
\int d^4x \,
A_\mu^{h,a} A_\mu^{h,a}
-
M^2
\int d^4x \,
\left(
\bar{\varphi}_\mu^{ab} \varphi_\mu^{ab}
-
\bar{\omega}_\mu^{ab} \omega_\mu^{ab}
\right),
\label{eq:condensates}
\end{align}
where $m^2$ and $M^2$ are dynamically generated mass scales associated with the
condensates $\langle A_\mu^{h,a} A_\mu^{h,a} \rangle$ and
$\langle \bar{\varphi}\varphi - \bar{\omega}\omega \rangle$, respectively.

Throughout this work, all quantities are defined in Euclidean space.
No assumption of a fundamental Minkowski formulation of the RGZ theory is made.
This is consistent with both the lattice formulation of YM theory and
the intrinsically nonperturbative nature of the observables under consideration.
In particular, this choice is deliberate, as glueball masses are Euclidean observables
customarily extracted from the long-distance behavior of Euclidean correlation functions.

Nonetheless, the central nonperturbative ingredient in the present analysis is the gluon
two-point function obtained from the RGZ framework.
In the BRST-invariant formulation based on the transverse field $A_\mu^h$,
the gluon propagator is well defined, multiplicatively renormalizable,
and directly comparable to lattice results in the Landau gauge.
At tree level, the RGZ action yields the following transverse gluon propagator
in four-dimensional Euclidean space-time~\cite{Dudal2008RGZ,Dudal2010LatticeFit}:
\begin{align}
D_{\mu\nu}^{ab}(p)
=
\delta^{ab}
\left(
\delta_{\mu\nu}
-
\frac{p_\mu p_\nu}{p^2}
\right)
\mathcal{D}(p^2),
\label{eq:RGZtensor}
\end{align}
with the scalar form factor
\begin{align}
\mathcal{D}(p^2)
=
\frac{p^2 + M^2}
{p^4 + (m^2 + M^2) p^2 + m^2 M^2 + 2 g^2 N \gamma^4}.
\label{eq:RGZscalar}
\end{align}
Although obtained at tree level, this propagator effectively resums dominant
infrared dynamics through the presence of the condensates and provides an accurate
description of lattice data over a broad momentum range.
The propagator~\eqref{eq:RGZscalar} exhibits a decoupling-type infrared behavior,
remaining finite at vanishing momentum,
\begin{align}
\mathcal{D}(0) = \frac{M^2}{m^2 M^2 + 2 g^2 N \gamma^4} \neq 0,
\end{align}
in qualitative and quantitative agreement with large-volume lattice simulations
of YM theory in the Landau gauge~\cite{CucchieriMendes2008,Bogolubsky2009,Maas2013}.

A distinctive feature of the RGZ propagator is its analytic structure.
The denominator of Eq.~\eqref{eq:RGZscalar} is a quartic polynomial in $p^2$,
which generically factorizes into two complex conjugate poles,
\begin{align}
p^2 = - m_\pm^2 , \qquad
m_\pm^2 = \mu^2 \pm i \, \theta^2 ,
\end{align}
with $\mu^2, \theta^2 > 0$.
As a consequence, the gluon propagator violates reflection positivity,
a hallmark of confinement~\cite{AguilarNataleRodriguesdaSilva2003, Dudal2008RGZ,CucchieriMendes2008}.
Such complex conjugate poles can be interpreted as an effective parametrization
of infrared gluonic correlations rather than physical asymptotic states, and they
provide a controlled analytic encoding of confinement effects.
Despite the absence of a standard K\"all\'en--Lehmann representation with a positive
spectral density at the level of elementary fields, the RGZ framework allows one to
consistently define and renormalize gauge-invariant composite operators, whose
correlation functions possess a well-defined analytic structure suitable for the
extraction of physical observables.
In the present work, the propagator~\eqref{eq:RGZtensor}--\eqref{eq:RGZscalar}
is taken as nonperturbative infrared input for the bound-state analysis.
The use of the RGZ gluon propagator thus provides a self-consistent and
lattice-compatible starting point for the formulation of a Euclidean
BSE describing glueball bound states.

\section{Glueball states and gauge-invariant composite operators}
\label{sec:operators}

Glueballs are color-singlet bound states generated by the self-interactions of YM fields. In a confining theory, they do not correspond to poles of
elementary gluon propagators, but rather appear as isolated poles, or dominant
exponential scales, in correlation functions of gauge-invariant composite
operators. This notion of the physical spectrum is standard in confining gauge
theories and underlies both lattice simulations and continuum approaches based
on operator correlation functions.

In Euclidean space, consider a local, gauge-invariant operator
$\mathcal{O}(x)$ carrying definite quantum numbers $J^{PC}$.
Its two-point correlation function is defined as
\begin{align}
\Pi(x-y)
=
\langle \mathcal{O}(x)\,\mathcal{O}(y)\rangle ,
\label{eq:Pi_xy_def}
\end{align}
which is translation invariant and thus depends only on $z=x-y$.
It admits the Fourier representation
\begin{align}
\Pi(p)
=
\int d^4 z\, e^{-ip\cdot z}\,\Pi(z)\,.
\label{eq:Pi_p_def}
\end{align}
If $\mathcal{O}$ couples to a discrete lowest-lying state of mass $M$ in the
given channel, the Euclidean correlator at large Euclidean time separation
behaves as
\begin{align}
\int d^3\bm{x}\;
\Pi(\tau,\bm{x})
\;\xrightarrow[\tau\to\infty]{}\;
\mathcal{Z}\,e^{-M\tau}\,,
\label{eq:large_tau_decay}
\end{align}
which is the standard lattice route to extracting glueball masses.
Equivalently, in momentum space one expects (up to subtractions) a spectral
representation of the form
\begin{align}
\Pi(p^2)
=
\int_{0}^{\infty} ds\;
\frac{\rho(s)}{s+p^2}\,,
\qquad
\rho(s)\ge 0,
\label{eq:KL_representation}
\end{align}
whenever $\mathcal{O}$ is gauge invariant and the corresponding physical Hilbert
space is positive.
In practice, approximate representations or truncations of $\Pi(p^2)$ are often
employed, provided they preserve a sensible spectral density in the physical
channels.

In this work we focus on the lightest scalar glueball channel,
\begin{align}
J^{PC} = 0^{++}\,.
\end{align}
The lowest-dimensional local gauge-invariant operator carrying these quantum
numbers is
\begin{align}
\mathcal{O}_{0^{++}}(x)
=
\frac{1}{4}\,F_{\mu\nu}^a(x)\,F_{\mu\nu}^a(x),
\label{eq:O_scalar_def}
\end{align}
where the YM field-strength tensor is
\begin{align}
F_{\mu\nu}^a
=
\partial_\mu A_\nu^a - \partial_\nu A_\mu^a
+ g f^{abc} A_\mu^b A_\nu^c\,.
\label{eq:Fmunu_def}
\end{align}
The normalization factor $\tfrac14$ in Eq.~\eqref{eq:O_scalar_def} is conventional
and plays no role in the bound-state condition.
Since $\mathcal{O}_{0^{++}}$ is gauge invariant, its correlator represents a
physical observable.
Within the BRST-invariant RGZ formulation based on
the transverse field $A_\mu^h$, this operator can be consistently renormalized,
and its correlation functions can be computed in a controlled way.
This point is essential, because the elementary gluon propagator in RGZ violates
reflection positivity and typically exhibits complex conjugate poles; therefore,
physical information must be extracted at the level of gauge-invariant composite
operators rather than from elementary fields.

For the purposes of a BSE treatment, it is instructive
to make explicit the leading ``two-gluon'' content of $\mathcal{O}_{0^{++}}$.
Expanding Eq.~\eqref{eq:Fmunu_def} in powers of the gauge field, we write
\begin{align}
F_{\mu\nu}^a = F_{\mu\nu}^{a,(1)} + F_{\mu\nu}^{a,(2)},
\end{align}
with
\begin{align}
F_{\mu\nu}^{a,(1)} &= \partial_\mu A_\nu^a - \partial_\nu A_\mu^a,
\\
F_{\mu\nu}^{a,(2)} &= g f^{abc} A_\mu^b A_\nu^c.
\end{align}
The scalar operator then decomposes as
\begin{align}
\mathcal{O}_{0^{++}}
=
\frac14\,F^{a,(1)}_{\mu\nu}F^{a,(1)}_{\mu\nu}
+\frac12\,F^{a,(1)}_{\mu\nu}F^{a,(2)}_{\mu\nu}
+\frac14\,F^{a,(2)}_{\mu\nu}F^{a,(2)}_{\mu\nu}.
\label{eq:O_expand}
\end{align}
The first term is quadratic in $A_\mu^a$ and provides the minimal two-gluon
component relevant for a BSE description.
In momentum space one has
\begin{align}
F_{\mu\nu}^{a,(1)}(p)
=
i\big(p_\mu A_\nu^a(p) - p_\nu A_\mu^a(p)\big),
\end{align}
leading to
\begin{align}
\frac14\,F^{a,(1)}_{\mu\nu}(p)\,F^{a,(1)}_{\mu\nu}(-p)
=
\frac12\,
A_\mu^a(p)\,
\Big(p^2 \delta_{\mu\nu} - p_\mu p_\nu\Big)\,
A_\nu^a(-p).
\label{eq:O_quadratic_momentum}
\end{align}
This expression explicitly shows that the scalar operator projects onto
transverse gluonic modes, as expected in the Landau gauge and consistent with
the tensor structure of the RGZ gluon propagator.
The remaining terms in Eq.~\eqref{eq:O_expand} are cubic and quartic in the gauge
field and, from the viewpoint of a bound-state truncation, correspond to higher
Fock components and interaction corrections.
In a BSE framework, such contributions are systematically incorporated through
the interaction kernel rather than being kept explicitly in the interpolating
operator.

The color structure of $\mathcal{O}_{0^{++}}$ is the trivial singlet.
At the level of the leading quadratic term~\eqref{eq:O_quadratic_momentum}, this
corresponds to contracting two adjoint indices with $\delta^{ab}$.
Accordingly, the color-singlet projector for a two-gluon state reads
\begin{align}
\mathcal{P}^{ab}_{\mathbf{1}}
=
\frac{\delta^{ab}}{\sqrt{N^2-1}},
\end{align}
up to normalization conventions.
Parity and charge-conjugation quantum numbers follow from the transformation
properties of $F_{\mu\nu}^aF_{\mu\nu}^a$.
In Minkowski space one has $F_{\mu\nu}F^{\mu\nu}=2(\bm{B}^2-\bm{E}^2)$, which is
even under both parity and charge conjugation.
In Euclidean space, the operator remains a scalar under $O(4)$ rotations and
assumes $P=+$ and $C=+$ from the fundamental Minkowski theory, thereby coupling
to $0^{++}$ states.

There are two complementary continuum strategies to access the glueball spectrum:
\begin{itemize}
\item[(i)]
Compute the correlator $\Pi(p^2)$ of a gauge-invariant operator, such as
$\mathcal{O}_{0^{++}}$, using an effective infrared description (here RGZ), and
extract the mass from its spectral representation or from suitable moment or
sum-rule techniques.
This approach is particularly natural within RGZ, where analytic expressions
for infrared YM correlators are available.
\item[(ii)]
Formulate a bound-state equation for an effective two-body amplitude
$\Gamma(p;P)$, whose constituents are infrared gluonic degrees of freedom
described by the RGZ propagator, and determine $P^2=-M^2$ from the existence of
nontrivial solutions.
\end{itemize}

In this work we follow route (ii), which gives direct access to the glueball
Bethe--Salpeter amplitude and enables for a systematic analysis of truncation and
kernel dependence, and at the same time remaining entirely within Euclidean signature.
The fundamental observation justifying a BSE formulation is that, at leading order in a
skeleton expansion, the two-point function of a bilinear gauge-invariant operator
receives contributions that can be organized into two-particle reducible gluon
ladders.
Resumming these ladders leads to an eigenvalue problem whose homogeneous form is
precisely a BSE.
Operationally, we treat the scalar glueball as a two-gluon bound state in the
color-singlet channel, with the RGZ gluon propagator as the constituent two-point
function and with an interaction kernel that captures the dominant infrared
gluonic exchanges.
Although the RGZ gluon propagator contains complex conjugate poles, this does
not obstruct the extraction of physical glueball masses.
The bound-state condition and mass determination are formulated entirely in
Euclidean space and anchored to gauge-invariant quantities, in close analogy
with lattice calculations.
The role of the RGZ propagator is to provide a realistic, infrared-saturated
two-point input consistent with confinement-related positivity violation, whereas
physical poles emerge only in composite, gauge-invariant channels.

With these points established, we now turn to the explicit formulation of the
Euclidean BSE for the $0^{++}$ glueball.
\section{Euclidean Bethe--Salpeter equation for the scalar glueball}
\label{sec:BSE}

In this section we formulate a Euclidean BSE for the
scalar glueball in the channel $J^{PC}=0^{++}$, using the RGZ gluon propagator as nonperturbative input as previously mentioned.
The construction follows standard bound-state techniques in continuum quantum
field theory and is adapted here to the case of confined gluonic degrees of
freedom in YM theory.

We start from the connected four-point Green function of gluon fields,
projected onto the color-singlet channel,
\begin{align}
G^{(4)}_{\mu\nu\rho\sigma}(x_1,x_2;x_3,x_4)
=
\big\langle
A_\mu^a(x_1)\, A_\nu^a(x_2)\,
A_\rho^b(x_3)\, A_\sigma^b(x_4)
\big\rangle_{\text{conn}},
\label{eq:G4_def}
\end{align}
where repeated color indices are summed over.
In momentum space, and after amputation of external legs, this object defines
the two-particle scattering kernel for gluonic correlations in the adjoint
representation.
If a bound state with total Euclidean momentum $P$ exists in the corresponding
channel, general field-theoretical arguments imply that the four-point function
develops a pole of the form
\begin{align}
G^{(4)} \;\sim\;
\frac{
\Gamma_{\mu\nu}(p;P)\,
\overline{\Gamma}_{\rho\sigma}(q;P)
}{
P^2 + M^2
}
\qquad
\text{for } P^2 \to -M^2,
\label{eq:4pt_pole}
\end{align}
in which $M$ is the bound-state mass and
$\Gamma_{\mu\nu}(p;P)$ denotes the Bethe--Salpeter amplitude,
that depends on the relative momentum $p$ and the total momentum $P$.
This pole structure follows from analyticity and completeness and does not rely
on the existence of asymptotic gluon states, which are absent in a confining
theory.

Resumming the two-particle reducible contributions in the $s$-channel leads to
a homogeneous integral equation for the Bethe--Salpeter amplitude,
\begin{align}
\Gamma_{\mu\nu}(p;P)
=
\int\!\frac{d^4 q}{(2\pi)^4}\;
\mathcal{K}_{\mu\nu\rho\sigma}(p,q;P)\,
D_{\rho\alpha}(q_+)\,
D_{\sigma\beta}(q_-)\,
\Gamma_{\alpha\beta}(q;P),
\label{eq:BSE_tensor}
\end{align}
where $q_\pm = q \pm P/2$ and
$D_{\mu\nu}$ is the RGZ gluon propagator introduced in
Eqs.~\eqref{eq:RGZtensor}--\eqref{eq:RGZscalar}.
The kernel $\mathcal{K}_{\mu\nu\rho\sigma}$ collects all two-particle irreducible
interactions in the chosen truncation.
In the present work we employ a symmetry-preserving ladder truncation,
corresponding to the lowest-order skeleton expansion.
This approximation captures the dominant infrared interactions responsible for
binding, whereas keeping the analysis analytically transparent.
Higher-order corrections, for instance crossed-ladder diagrams and explicit three-
and four-gluon vertex dressings, are neglected at this stage and will be
addressed in future refinements.

For the scalar glueball channel $J^{PC}=0^{++}$, the Bethe--Salpeter amplitude can
be decomposed in terms of transverse tensors consistent with Lorentz symmetry,
parity, and charge conjugation.
At leading order, only one independent scalar structure contributes, and we
write
\begin{align}
\Gamma_{\mu\nu}(p;P)
=
\left(
\delta_{\mu\nu}
-
\frac{P_\mu P_\nu}{P^2}
\right)
\Phi(p;P),
\label{eq:Gamma_decomp}
\end{align}
in which $\Phi(p;P)$ is a scalar Bethe--Salpeter amplitude.
Substituting the decomposition~\eqref{eq:Gamma_decomp} into
Eq.~\eqref{eq:BSE_tensor} and performing the tensor contractions yields a scalar
integral equation of the form
\begin{align}
\Phi(p;P)
=
\int\!\frac{d^4 q}{(2\pi)^4}\;
\mathcal{K}(p,q;P)\,
\mathcal{D}(q_+^2)\,
\mathcal{D}(q_-^2)\,
\Phi(q;P),
\label{eq:BSE_scalar}
\end{align}
where $\mathcal{D}(p^2)$ is the scalar part of the RGZ gluon propagator and
$\mathcal{K}(p,q;P)$ denotes an effective scalar kernel determined by the chosen
truncation.
Equation~\eqref{eq:BSE_scalar} constitutes a homogeneous eigenvalue problem.
Nontrivial solutions exist only for discrete values of $P^2$, and the lowest
eigenvalue defines the mass of the lightest scalar glueball,
\begin{align}
P^2 = -M_{0^{++}}^2.
\end{align}
A potential concern in the RGZ framework, however, is the presence of complex conjugate
poles in the gluon propagator.
Indeed, this feature does not invalidate the Bethe--Salpeter construction for
several reasons:
\begin{itemize}
\item
The BSE is formulated entirely in Euclidean space, where the RGZ propagator is
analytic and free of singularities on the integration domain.
\item
The bound-state condition is imposed on a gauge-invariant composite amplitude
and does not correspond to a pole of the elementary gluon propagator.
\item
Glueball masses are extracted from Euclidean correlation functions, in direct
analogy with lattice simulations of YM theory.
\end{itemize}
The complex analytic structure of the RGZ propagator should therefore be
interpreted as a manifestation of confinement at the level of elementary
fields, whereas physical poles emerge only in composite, gauge-invariant channels.

With the Euclidean BSE established, we now turn to its
numerical solution, renormalization procedure, and stability analysis.

\section{Numerical strategy and renormalization}
\label{sec:numerics}

In this section we detail the numerical reduction of the Euclidean BSE derived in Sec.~\ref{sec:BSE}, and clarify the
role of renormalization when the RGZ gluon
propagator is employed as an infrared-complete effective input.
All calculations are performed in Euclidean space.
The total momentum $P$ of the bound state satisfies the on-shell condition
\begin{align}
P^2 = - M_{0^{++}}^2 < 0,
\end{align}
while $p$ denotes the relative momentum.
We introduce the shifted loop momenta
\begin{align}
q_\pm = q \pm \frac{P}{2}.
\end{align}
Owing to $O(4)$ covariance, the scalar Bethe--Salpeter amplitude depends only on
the invariants
\begin{align}
p^2, \qquad
z_p := \frac{p\cdot P}{\sqrt{p^2 P^2}}, \qquad
P^2,
\end{align}
and analogously for $q$.
We choose a frame where the total momentum points along the Euclidean 4-axis,
\begin{align}
P_\mu = (0,0,0,\sqrt{P^2}),
\end{align}
so that $z_p\in[-1,1]$ and
\begin{align}
q_\pm^2
=
q^2 + \frac{P^2}{4} \pm \sqrt{q^2 P^2}\, z_q .
\label{eq:qpm2}
\end{align}
All integration variables remain real for $P^2>0$ in this convention.
Although the RGZ gluon propagator $\mathcal{D}(k^2)$ possesses complex
conjugate poles in the complex $k^2$ plane, the Euclidean integration domain
never crosses these singularities.

Using the tensor decomposition discussed in Sec.~\ref{sec:BSE}, the scalar BSE
can be written as
\begin{align}
\Phi(p^2,z_p;P^2)
=
\int\!\frac{d^4 q}{(2\pi)^4}\;
\mathcal{K}(p,q;P)\,
\mathcal{D}(q_+^2)\,
\mathcal{D}(q_-^2)\,
\Phi(q^2,z_q;P^2),
\label{eq:BSE_scalar_recalled}
\end{align}
where $\mathcal{K}$ denotes the effective scalar kernel resulting from color and
tensor contractions.
For numerical purposes it is convenient to cast the BSE into an eigenvalue
problem by introducing a parameter $\lambda(P^2)$,
\begin{align}
\Phi(p)
=
\lambda(P^2)
\int\!\frac{d^4 q}{(2\pi)^4}\;
\mathcal{K}(p,q;P)\,
\mathcal{D}(q_+^2)\,
\mathcal{D}(q_-^2)\,
\Phi(q),
\label{eq:eigenvalue_form}
\end{align}
where $\Phi(p)\equiv\Phi(p^2,z_p;P^2)$.
For a fixed kernel model and fixed RGZ parameters, the bound-state mass is
obtained from the condition
\begin{align}
\lambda_{\max}(P^2) = 1,
\end{align}
where $\lambda_{\max}(P^2)$ denotes the largest eigenvalue of the integral
operator.
In practice, $\lambda_{\max}(P^2)$ is monotonic for a wide class of ladder-type
kernels, enabling for a stable one-dimensional root-finding procedure.
This strategy is standard in functional approaches to bound states based on
DSEs and BSEs
\cite{AlkoferVonSmekal2001,Huber2020}.

The angular dependence of the amplitude is expanded in Chebyshev polynomials,
\begin{align}
\Phi(p^2,z_p;P^2)
=
\sum_{n=0}^{N_z} \Phi_n(p^2;P^2)\, T_n(z_p),
\label{eq:cheb_expand}
\end{align}
with $T_n(z)=\cos(n\arccos z)$.
The truncation order $N_z$ is chosen such that the extracted mass is stable under
increasing $N_z$.
The four-dimensional integration measure is written as
\begin{align}
\int \frac{d^4 q}{(2\pi)^4}
=
\frac{1}{(2\pi)^4}
\int_0^\infty dq\, q^3
\int d\Omega_3,
\end{align}
with the angular dependence parameterized by $z_q$.
For the radial integration we employ the mapping
\begin{align}
q^2 = \Lambda^2 \frac{x}{1-x},\qquad x\in[0,1),
\label{eq:map_radial}
\end{align}
which yields
\begin{align}
q^3 dq
=
\frac{\Lambda^4}{2}\,
\frac{x\,dx}{(1-x)^3}.
\label{eq:jacobian_q3dq}
\end{align}
The scale $\Lambda$ controls the transition between infrared and ultraviolet
sampling and is varied to ensure numerical stability.

After discretization, Eq.~\eqref{eq:eigenvalue_form} becomes a matrix eigenvalue
problem,
\begin{align}
\boldsymbol{\Phi} = \lambda(P^2)\,\mathbf{M}(P^2)\,\boldsymbol{\Phi},
\end{align}
where $\mathbf{M}(P^2)$ encodes the kernel, propagators and quadrature weights.
For each trial value of $P^2$, the largest eigenvalue is computed using standard
iterative methods.
The homogeneous BSE determines the amplitude only up to an overall
normalization.
A physical normalization can be imposed through the canonical condition derived
from the residue of the four-point function at the bound-state pole,
\begin{align}
1
=
\left.
\frac{d}{dP^2}
\int\!\frac{d^4 q}{(2\pi)^4}
\frac{d^4 k}{(2\pi)^4}\;
\overline{\Gamma}(q;P)\,
D(q_+)D(q_-)\,
\Gamma(k;P)
\right|_{P^2=-M^2}.
\end{align}
This normalization is essential for computing decay constants or transition
form factors, but it is not required for the extraction of the bound-state mass
itself.

A fundamental point of our present approach concerns renormalization.
We do not renormalize the BSE in the perturbative FP sense, since we
do not employ bare propagators or vertices.
Instead, the RGZ gluon propagator is taken as a renormalized effective input,
whose parameters $(m^2,M^2,\gamma^4)$ are fixed by RGZ gap equations and/or
lattice fits in a given renormalization scheme
\cite{Dudal2008RGZ,Dudal2010LatticeFit}.
Any multiplicative wave-function renormalization is already absorbed into these
parameters.

The effective interaction strength entering the kernel is treated as a model
parameter.
Varying it within a controlled range provides an estimate of truncation
uncertainties, in direct analogy with common practice in DSEs and
BSEs studies.
Ultraviolet stability is ensured by the $1/p^2$ falloff of the RGZ propagator at
large momenta, and explicit cutoff dependence is monitored numerically.

It is worth emphasizing that the present strategy is conceptually close to the
infrared moment problem approach of
Ref.~\cite{DudalGuimaraesSorella2011PRL}.
In that framework, glueball mass estimates display a residual dependence on the
subtraction scale $T$, which is not a physical parameter and is fixed by an
optimization criterion such as minimal sensitivity.
In the present Euclidean BSE formulation, an analogous role is played by the
effective kernel strength (and, to a lesser extent, numerical resolution
parameters), whose variation provides a controlled estimate of systematic
uncertainties associated with the truncation.
Physical predictions are extracted from regions where the resulting masses
exhibit maximal stability under such variations.

Finally, the robustness of the numerical results is assessed through:
\begin{enumerate}
\item[(i)] convergence with respect to the number of radial and angular grid points,
$(N_q,N_z)$;
\item[(ii)] stability under variations of the mapping scale $\Lambda$ introduced in
Eq.~\eqref{eq:map_radial};
\item[(iii)] independence of the extracted mass under changes of the ultraviolet
cutoff;
\item[(iv)] variation of the effective kernel strength to quantify truncation
uncertainties;
\item[(v)] propagation of uncertainties associated with the infrared RGZ parameters.
\end{enumerate}
Together, these tests define a systematic uncertainty band for the extracted
scalar glueball mass.

\section{Results and comparison with RGZ correlator approaches}
\label{sec:results}

In this section we present the numerical solutions of the Euclidean BSE derived in the previous sections and compare the resulting scalar
glueball mass with estimates obtained from correlator-based approaches within the
RGZ framework.

At first sight, the numerical strategy adopted here differs from the correlator-based
infrared moment analysis of
Refs.~\cite{DudalGuimaraesSorella2011PRL,Dudal2010LatticeFit}, which displays the glueball
masses as functions of the subtraction scale $T$.
This difference is, however, purely methodological.
In the correlator approach, $T$ is a non-physical parameter introduced by the subtraction
procedure, and physical masses are extracted from regions of minimal sensitivity.
In the present Euclidean Bethe--Salpeter formulation, an analogous role is played by the
effective kernel strength $g_C^2$, whose variation probes the truncation dependence of the
bound-state equation.
Although the graphical representations are different, both approaches rely on the same
RGZ gluon propagator as infrared input and yield consistent scalar glueball mass estimates
in the vicinity of $2~\mathrm{GeV}$.

To render the integral equation numerically tractable, we adopt a minimal truncation
in which the Bethe--Salpeter kernel is approximated by an effective constant strength
$g_C^2$. In addition, we restrict ourselves to the dominant $s$--wave component of
the amplitude by retaining only the lowest Chebyshev moment,
$\Phi(p^2) \equiv \Phi_0(p^2)$, and neglecting its explicit angular dependence.
Within these assumptions, the eigenvalue equation~\eqref{eq:eigenvalue_form} reduces
to a one-dimensional integral equation,
\begin{align}
\lambda(P^2)\,\Phi(p^2)
&=
g_C^2
\int_0^\infty\!dq^2\,q^2
\int_{-1}^{1}\!dz\,
\mathcal{D}(q_+^2)\,
\mathcal{D}(q_-^2)\,
\Phi(q^2),
\label{eq:simplified_BSE_scalar}
\end{align}
where the shifted momenta $q_\pm^2$ are given by Eq.~\eqref{eq:qpm2} and the integration
over relative angles has been reduced to a single variable $z$.
The integrals are discretized on finite domains using Gauss--Legendre quadratures
for both $q^2$ and $z$.

The gluon propagator $\mathcal{D}(p^2)$ entering
Eq.~\eqref{eq:simplified_BSE_scalar} is the RGZ propagator given in
Eq.~\eqref{eq:RGZscalar}, evaluated with representative parameter values motivated
by lattice fits.
In particular, Ref.~\cite{Dudal2010LatticeFit} reports
$m^2 + M^2 \simeq 0.34~\mathrm{GeV}^2$, $M^2 \simeq 2.15~\mathrm{GeV}^2$ and
$m^2 M^2 + 2 g^2 N \gamma^4 \simeq 0.26~\mathrm{GeV}^4$.
For the numerical analysis presented here we adopt the values
$m^2 = 0.5~\mathrm{GeV}^2$, $M^2 = 2.0~\mathrm{GeV}^2$ and
$\lambda_4 = 0.26~\mathrm{GeV}^4$, which lie within the phenomenologically relevant
window.

\begin{figure}[t]
\centering
\includegraphics[width=0.6\linewidth]{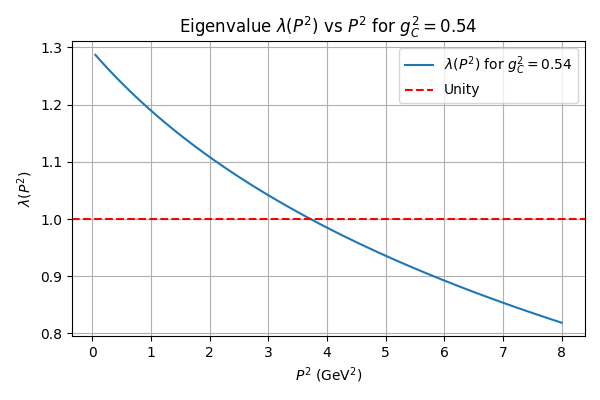}
\caption{Largest eigenvalue $\lambda(P^2)$ of the Bethe--Salpeter kernel as a function
of $P^2$ for an effective coupling $g_C^2 = 0.54$.
The intersection with the horizontal line $\lambda=1$ fixes the scalar glueball
mass.}
\label{fig:lambda}
\end{figure}

For fixed RGZ parameters, Eq.~\eqref{eq:simplified_BSE_scalar} is solved by varying
the total momentum $P^2$ and determining the largest eigenvalue $\lambda(P^2)$.
The bound--state condition corresponds to
$\lambda(P^2) = 1$, with the scalar glueball mass given by
$M_{0^{++}}^2 = -P^2$.
Since $\lambda(P^2)$ scales linearly with the effective coupling $g_C^2$, one may
equivalently view the procedure as tuning $g_C^2$ until a crossing occurs.
Figure~\ref{fig:lambda} shows a representative example of the eigenvalue
$\lambda(P^2)$ as a function of $P^2$ for $g_C^2 = 0.54$.
The eigenvalue decreases monotonically with increasing $P^2$, and the crossing
with $\lambda=1$ unambiguously determines the bound--state mass.

The resulting scalar glueball masses obtained for several values of $g_C^2$ are
listed in Table~\ref{tab:g2C_mass}.
For couplings in the range $g_C^2 \simeq 0.52$--$0.58$, the extracted masses lie
between approximately $1.7$ and $2.3~\mathrm{GeV}$, with a preferred value close to
$M_{0^{++}} \simeq 1.9~\mathrm{GeV}$ around $g_C^2 \simeq 0.54$.

\begin{table}[t]
\caption{Scalar glueball mass $M_{0^{++}}$ obtained from the simplified Euclidean
Bethe--Salpeter equation for several values of the effective coupling $g_C^2$.
The RGZ parameters $(m^2,M^2,\lambda_4)$ are kept fixed as described in the text.}
\label{tab:g2C_mass}
\begin{tabular}{c c}
\hline\hline
$g_C^2$ & $M_{0^{++}}$ (GeV) \\
\hline
0.50 & 1.55 \\
0.52 & 1.75 \\
0.54 & 1.93 \\
0.56 & 2.10 \\
0.58 & 2.26 \\
0.60 & 2.41 \\
\hline\hline
\end{tabular}
\end{table}
These values are in remarkable agreement with the scalar glueball mass obtained
from RGZ correlator-based analyses.
In particular, the infrared moment problem studied in
Refs.~\cite{DudalGuimaraesSorella2011PRL,Dudal2010LatticeFit}
yields an estimate $M_{0^{++}} \approx 1.96~\mathrm{GeV}$ for $\mathrm{SU}(3)$,
using the same RGZ gluon propagator as infrared input.
In that approach, the two--point function of the gauge--invariant operator
$\mathcal{O}_{0^{++}} = \tfrac{1}{4} F_{\mu\nu}^a F_{\mu\nu}^a$ is expanded in infrared
moments and analyzed via Pad\'e approximants, with confinement encoded through the
complex conjugate poles (``$i$--particles'') of the RGZ propagator.

The present Bethe--Salpeter analysis offers a complementary perspective.
While the correlator method directly probes the analytic structure of gauge--invariant
two--point functions, the BSE framework interprets the scalar glueball as a bound state
of infrared gluonic degrees of freedom and provides access to its Bethe--Salpeter
amplitude.
Despite the simplicity of the kernel employed here, both approaches lead to
consistent mass estimates, underscoring the robustness of the RGZ description of
infrared YM dynamics.
A more refined treatment, including momentum--dependent kernels and higher
Chebyshev moments, would allow for a quantitative assessment of systematic
uncertainties and enable the extension of the present analysis to other glueball
channels.

\section{Conclusions and outlook}
\label{sec:conclusions}

In this work we have formulated and numerically solved a Euclidean BSE for the
lightest scalar glueball in pure YM theory, using the RGZ gluon propagator as an
infrared-complete input.
Within a minimal truncation characterized by an effective constant kernel strength
and the dominant $s$--wave amplitude, we extracted scalar glueball masses in the
range $M_{0^{++}}\simeq 1.7$--$2.3~\mathrm{GeV}$ for representative values of the
kernel parameter $g_C^2$.
A preferred value around $1.9~\mathrm{GeV}$ was obtained for $g_C^2\simeq 0.54$,
in close agreement with correlator-based estimates within the RGZ framework
\cite{DudalGuimaraesSorella2011PRL,Dudal2010LatticeFit} and with contemporary lattice
determinations.
This concordance reinforces the robustness of the RGZ description of infrared
gluodynamics and provides a nontrivial cross-check between bound-state and
correlator approaches.

Our analysis showd that the BSE framework, even in a simplified truncation,
is capable of capturing essential features of glueball dynamics when supplied with
realistic nonperturbative propagators.
We emphasized the conceptual correspondence between the effective kernel strength
in the BSE and the subtraction-scale dependence encountered in infrared moment
problems, showing that both play the role of nonphysical control parameters whose
systematic variation quantifies truncation uncertainties.
In this sense, the RGZ gluon propagator, with its complex conjugate poles encoding
confinement, serves as a common infrared input across complementary methodologies.

Several directions for future investigations naturally emerge from the present
study.
First, the inclusion of momentum-dependent kernels and higher Chebyshev moments
would allow for a more refined assessment of systematic uncertainties and may enable
the resolution of excited glueball states.
Such improvements would bring the present analysis closer to the level of
sophistication achieved in meson and baryon studies within DSE/BSE frameworks.

A particularly promising avenue is the integration of our recent advances
in physics-informed
neural networks (PINNs) for solving DSEs~\cite{Terin2025PINNsDSE,Terin2025SpectralMinkowski}into the present BSE approach.
PINNs provide an efficient strategy for tackling nonlinear integral equations by
embedding physical constraints directly into the learning architecture.
Coupling PINN-based DSE solutions with a BSE treatment of glueball bound states could
yield a powerful and flexible computational framework for exploring nonperturbative
YM dynamics on multiple quantum-number channels.
Another perspective is opened by our recent progress toward a unified
formulation of gauge fixing in the presence of Gribov copies, interpolating between
the ST and RGZ approaches~\cite{Terin2025UnifiedST_RGZ}.
By combining the present BSE program within such a unified gauge-fixing theory may
give novel information into the interplay between gauge copies, infrared dynamics
and bound-state formation.

Altogether, these developments point toward a comprehensive continuum approach to
YM spectroscopy that harmonizes infrared analytic inputs, numerical bound-state
techniques and modern computational tools.
We believe that such an integrated program will significantly advance the
understanding of pure gauge bound states and pave the way for future applications
in full QCD.


\bibliographystyle{unsrt}
\bibliography{RGZ_BSE}

@article{YangMills1954,
  author       = {Yang, C. N. and Mills, R. L.},
  title        = {Conservation of Isotopic Spin and Isotopic Gauge Invariance},
  journal      = {Phys. Rev.},
  volume       = {96},
  pages        = {191--195},
  year         = {1954},
  doi          = {10.1103/PhysRev.96.191}
}

@article{FaddeevPopov1967,
  author       = {Faddeev, L. D. and Popov, V. N.},
  title        = {Feynman Diagrams for the Yang--Mills Field},
  journal      = {Phys. Lett. B},
  volume       = {25},
  pages        = {29--30},
  year         = {1967},
  doi          = {10.1016/0370-2693(67)90067-6}
}

@article{Gribov1978,
  author       = {Gribov, V. N.},
  title        = {Quantization of Nonabelian Gauge Theories},
  journal      = {Nucl. Phys. B},
  volume       = {139},
  pages        = {1--19},
  year         = {1978},
  doi          = {10.1016/0550-3213(78)90175-X}
}

@article{Singer1978,
  author       = {Singer, I. M.},
  title        = {Some Remarks on the Gribov Ambiguity},
  journal      = {Commun. Math. Phys.},
  volume       = {60},
  pages        = {7--12},
  year         = {1978},
  doi          = {10.1007/BF01609471}
}

@article{Zwanziger1989,
  author       = {Zwanziger, D.},
  title        = {Local and Renormalizable Action from the Gribov Horizon},
  journal      = {Nucl. Phys. B},
  volume       = {323},
  pages        = {513--544},
  year         = {1989},
  doi          = {10.1016/0550-3213(89)90122-3}
}

@article{Zwanziger1990,
  author       = {Zwanziger, D.},
  title        = {Quantization of Gauge Fields, Classical Gauge Invariance, and Gluon Confinement},
  journal      = {Nucl. Phys. B},
  volume       = {345},
  pages        = {461--482},
  year         = {1990},
  doi          = {10.1016/0550-3213(90)90369-P}
}

@article{Dudal2008RGZ,
  author       = {Dudal, D. and Gracey, J. A. and Sorella, S. P. and Vandersickel, N. and Verschelde, H.},
  title        = {Refining the Gribov--Zwanziger approach in the Landau gauge: Infrared propagators in harmony with the lattice results},
  journal      = {Phys. Rev. D},
  volume       = {78},
  pages        = {065047},
  year         = {2008},
  doi          = {10.1103/PhysRevD.78.065047},
  eprint       = {0806.4348},
  archivePrefix= {arXiv},
  primaryClass = {hep-th}
}

@article{Dudal2010LatticeFit,
  author       = {Dudal, D. and Oliveira, O. and Vandersickel, N.},
  title        = {Indirect lattice evidence for the Refined Gribov--Zwanziger formalism and the gluon condensate $\langle A^2\rangle$ in the Landau gauge},
  journal      = {Phys. Rev. D},
  volume       = {81},
  pages        = {074505},
  year         = {2010},
  doi          = {10.1103/PhysRevD.81.074505},
  eprint       = {1002.2374},
  archivePrefix= {arXiv},
  primaryClass = {hep-lat}
}

@article{CucchieriMendes2008,
  author       = {Cucchieri, A. and Mendes, T.},
  title        = {Constraints on the IR behaviour of the gluon propagator in Yang--Mills theories},
  journal      = {Phys. Rev. Lett.},
  volume       = {100},
  pages        = {241601},
  year         = {2008},
  doi          = {10.1103/PhysRevLett.100.241601},
  eprint       = {0712.3517},
  archivePrefix= {arXiv},
  primaryClass = {hep-lat}
}

@article{Bogolubsky2009,
  author       = {Bogolubsky, I. L. and Ilgenfritz, E.-M. and M{\"u}ller-Preussker, M. and Sternbeck, A.},
  title        = {The Landau gauge gluon and ghost propagators in 4D {SU}(3) lattice gauge theory: Further results},
  journal      = {Phys. Lett. B},
  volume       = {676},
  pages        = {69--73},
  year         = {2009},
  doi          = {10.1016/j.physletb.2009.04.076},
  eprint       = {0901.0736},
  archivePrefix= {arXiv},
  primaryClass = {hep-lat}
}

@article{Boucaud2012,
  author       = {Boucaud, P. and Leroy, J. P. and Yaouanc, A. L. and Micheli, J. and Pene, O. and Rodriguez-Quintero, J.},
  title        = {The Infrared Behaviour of the Pure {Yang-Mills} Green Functions},
  journal      = {Few Body Syst.},
  volume       = {53},
  pages        = {387--436},
  year         = {2012},
  doi          = {10.1007/s00601-012-0434-4},
  eprint       = {1109.1936},
  archivePrefix= {arXiv},
  primaryClass = {hep-ph}
}

@article{Maas2013,
  author       = {Maas, A.},
  title        = {Describing gauge bosons at zero and finite temperature},
  journal      = {Phys. Rept.},
  volume       = {524},
  pages        = {203--300},
  year         = {2013},
  doi          = {10.1016/j.physrep.2012.11.002},
  eprint       = {1106.3942},
  archivePrefix= {arXiv},
  primaryClass = {hep-ph}
}

@article{SerreauTissier2012,
  author       = {Serreau, J. and Tissier, M.},
  title        = {Lifting the Gribov ambiguity in {Yang--Mills} theories},
  journal      = {Phys. Lett. B},
  volume       = {712},
  pages        = {97--103},
  year         = {2012},
  doi          = {10.1016/j.physletb.2012.04.041},
  eprint       = {1202.3432},
  archivePrefix= {arXiv},
  primaryClass = {hep-th}
}

@article{SerreauTissierTresmontant2015,
  author       = {Serreau, J. and Tissier, M. and Tresmontant, A.},
  title        = {Covariant gauges without Gribov copies},
  journal      = {Phys. Rev. D},
  volume       = {92},
  pages        = {105003},
  year         = {2015},
  doi          = {10.1103/PhysRevD.92.105003},
  eprint       = {1505.07270},
  archivePrefix= {arXiv},
  primaryClass = {hep-th}
}

@article{ReinosaSerreauTerinTissier2021,
  author       = {Reinosa, U. and Serreau, J. and Terin, R. Carmo and Tissier, M.},
  title        = {Symmetry restoration and the gluon mass in the Landau gauge},
  journal      = {SciPost Phys.},
  volume       = {10},
  number       = {2},
  pages        = {035},
  year         = {2021},
  doi          = {10.21468/SciPostPhys.10.2.035},
  eprint       = {2004.12413},
  archivePrefix= {arXiv},
  primaryClass = {hep-th}
}

@article{Sorella2015Nielsen,
  author       = {Sorella, S. P. and others},
  title        = {Nonperturbative aspects of Euclidean {Yang--Mills} theories in linear covariant gauges: {N}ielsen identities and a {BRST}-invariant two-point correlation function},
  journal      = {Phys. Rev. D},
  volume       = {92},
  pages        = {045039},
  year         = {2015},
  doi          = {10.1103/PhysRevD.92.045039},
  eprint       = {1503.05455},
  archivePrefix= {arXiv},
  primaryClass = {hep-th}
}

@article{Terin2025PINNsDSE,
  author       = {Terin, Rodrigo Carmo},
  title        = {Physics-informed neural networks viewpoint for solving the Dyson--Schwinger equations of quantum electrodynamics},
  journal      = {SciPost Phys. Core},
  volume       = {8},
  pages        = {054},
  year         = {2025},
  doi          = {10.21468/SciPostPhysCore.8.3.054},
  eprint       = {2411.02177},
  archivePrefix= {arXiv},
  primaryClass = {hep-ph}
}

@article{MeyersSwanson2012Glueballs,
  author       = {Meyers, J. and Swanson, E. S.},
  title        = {Spin-zero glueballs in the {Bethe--Salpeter} formalism},
  journal      = {Phys. Rev. D},
  volume       = {87},
  number       = {3},
  pages        = {036009},
  year         = {2013},
  doi          = {10.1103/PhysRevD.87.036009},
  eprint       = {1212.0058},
  archivePrefix= {arXiv},
  primaryClass = {hep-ph}
}

@article{SanchisAlepuz2015Glueballs,
  author       = {Sanchis-Alepuz, H. and Williams, R. and Alkofer, R. and Fischer, C. S.},
  title        = {On the gluon bound state spectrum in {Yang--Mills} theory},
  journal      = {Phys. Rev. D},
  volume       = {92},
  number       = {3},
  pages        = {034001},
  year         = {2015},
  doi          = {10.1103/PhysRevD.92.034001},
  eprint       = {1503.05896},
  archivePrefix= {arXiv},
  primaryClass = {hep-ph}
}

@article{HuberFischerSanchisAlepuz2020,
  author       = {Huber, M. Q. and Fischer, C. S. and Sanchis-Alepuz, H.},
  title        = {Glueball properties from the {Dyson--Schwinger}/{Bethe--Salpeter} equations},
  journal      = {Eur. Phys. J. C},
  volume       = {80},
  number       = {11},
  pages        = {1077},
  year         = {2020},
  doi          = {10.1140/epjc/s10052-020-08612-9},
  eprint       = {2004.00415},
  archivePrefix= {arXiv},
  primaryClass = {hep-ph}
}

@article{Terin2025SpectralMinkowski,
  author       = {Terin, Rodrigo Carmo},
  title        = {Spectral functions in Minkowski quantum electrodynamics from neural reconstruction: Benchmarking against dispersive Dyson--Schwinger integral equations},
  journal      = {arXiv},
  year         = {2025},
  eprint       = {2510.24728},
  archivePrefix= {arXiv},
  primaryClass = {hep-ph}
}

@article{AguilarNataleRodriguesdaSilva2003,
  author       = {Aguilar, A. C. and Natale, A. A. and Rodrigues da Silva, P. S.},
  title        = {Relating a gluon mass scale to an infrared fixed point in pure gauge QCD},
  journal      = {Phys. Rev. Lett.},
  volume       = {90},
  pages        = {152001},
  year         = {2003},
  doi          = {10.1103/PhysRevLett.90.152001},
  eprint       = {hep-ph/0212105},
  archivePrefix= {arXiv},
  primaryClass = {hep-ph}
}

@article{Sorella2017BRSTinvariant,
  author       = {Sorella, S. P. and others},
  title        = {A local and {BRST}-invariant {Yang--Mills} theory within the Gribov horizon},
  journal      = {Phys. Rev. D},
  volume       = {95},
  pages        = {045011},
  year         = {2017},
  doi          = {10.1103/PhysRevD.95.045011},
  eprint       = {1610.01590},
  archivePrefix= {arXiv},
  primaryClass = {hep-th}
}

@article{Terin2021FermionicHorizon,
  author       = {Terin, R. Carmo and others},
  title        = {All-order renormalizable refined {Gribov--Zwanziger} model with {BRST}-invariant fermionic horizon function in linear covariant gauges},
  journal      = {Phys. Rev. D},
  volume       = {104},
  number       = {5},
  pages        = {054048},
  year         = {2021},
  doi          = {10.1103/PhysRevD.104.054048},
  eprint       = {2104.08464},
  archivePrefix= {arXiv},
  primaryClass = {hep-th}
}

@article{DudalGuimaraesSorella2011PRL,
  author       = {Dudal, D. and Guimaraes, M. S. and Sorella, S. P.},
  title        = {Glueball Masses from an Infrared Moment Problem},
  journal      = {Phys. Rev. Lett.},
  volume       = {106},
  pages        = {062003},
  year         = {2011},
  doi          = {10.1103/PhysRevLett.106.062003},
  eprint       = {1010.3638},
  archivePrefix= {arXiv},
  primaryClass = {hep-th}
}

@article{AlkoferVonSmekal2001,
  author       = {Alkofer, R. and von Smekal, L.},
  title        = {The infrared behavior of {QCD} {Green}'s functions: Confinement, dynamical symmetry breaking, and hadrons as relativistic bound states},
  journal      = {Phys. Rept.},
  volume       = {353},
  pages        = {281--465},
  year         = {2001},
  doi          = {10.1016/S0370-1573(01)00010-2},
  eprint       = {hep-ph/0007355},
  archivePrefix= {arXiv}
}

@article{Huber2020,
  author       = {Huber, M. Q.},
  title        = {Nonperturbative properties of {Yang--Mills} theories},
  journal      = {Phys. Rept.},
  volume       = {879},
  pages        = {1--92},
  year         = {2020},
  doi          = {10.1016/j.physrep.2020.04.004},
  eprint       = {2002.04057},
  archivePrefix= {arXiv},
  primaryClass = {hep-ph}
}

@article{Pawlowski2007,
  author       = {Pawlowski, J. M.},
  title        = {Aspects of the functional renormalisation group},
  journal      = {Annals Phys.},
  volume       = {322},
  pages        = {2831--2915},
  year         = {2007},
  doi          = {10.1016/j.aop.2007.01.007},
  eprint       = {hep-th/0512261},
  archivePrefix= {arXiv}
}

@book{Nakanishi1971,
  author       = {Nakanishi, N.},
  title        = {Graph Theory and Feynman Integrals},
  publisher    = {Gordon and Breach},
  address      = {New York},
  year         = {1971}
}

@article{BetheSalpeter1951,
  author       = {Bethe, H. A. and Salpeter, E. E.},
  title        = {A Relativistic Equation for Bound-State Problems},
  journal      = {Phys. Rev.},
  volume       = {84},
  pages        = {1232--1242},
  year         = {1951},
  doi          = {10.1103/PhysRev.84.1232}
}

@article{CarbonellKarmanov2006,
  author       = {Carbonell, J. and Karmanov, V. A.},
  title        = {Solving Bethe--Salpeter equation in Minkowski space},
  journal      = {Eur. Phys. J. A},
  volume       = {27},
  pages        = {1--9},
  year         = {2006},
  doi          = {10.1140/epja/i2006-10013-9},
  eprint       = {hep-th/0601180},
  archivePrefix= {arXiv}
}

@article{CarbonellKarmanov2010,
  author       = {Carbonell, J. and Karmanov, V. A.},
  title        = {Solving Bethe--Salpeter equation for two fermions in Minkowski space},
  journal      = {Eur. Phys. J. A},
  volume       = {46},
  pages        = {387--397},
  year         = {2010},
  doi          = {10.1140/epja/i2010-11068-1},
  eprint       = {0912.0741},
  archivePrefix= {arXiv},
  primaryClass = {hep-ph}
}

@article{dePaulaFrederico2026,
  author       = {de Paula, Wayne and Frederico, Tobias},
  title        = {Minkowski Space Dynamics and Light-Front Projection},
  year         = {2026},
  eprint       = {2601.11760},
  archivePrefix= {arXiv},
  primaryClass = {hep-ph},
  note         = {arXiv preprint}
}

@article{Terin2025UnifiedST_RGZ,
  author       = {Terin, Rodrigo Carmo},
  title        = {Towards a unified viewpoint of Gribov--Zwanziger and Serreau--Tissier gauge fixing},
  year         = {2025},
  eprint       = {2510.18443},
  archivePrefix= {arXiv},
  primaryClass = {hep-th},
  note         = {arXiv preprint}
}

\end{document}